\documentclass[doublecol,figures]{epl2} 
\usepackage{graphicx}
\usepackage{amssymb,amsmath,amsthm,latexsym,rotating,setspace,placeins}
\usepackage[utf8]{inputenc}

\usepackage{xcolor,soul}

\begin{document} 

\title{Emergence of a collective crystal in a classical system with long-range interactions} 

\author{Alessio Turchi\inst{1}, Duccio Fanelli\inst{2} and Xavier Leoncini\inst{3}}
\institute{\inst{1}INAF Osservatorio Astrofisico di Arcetri, Largo Enrico Fermi 5, I-50125 Firenze, Italy\\
\inst{2}Dipartimento di Fisica e Astronomia, Universit\'a di Firenze and INFN and CSDC,s via Sansone, I-50019 Sesto Fiorentino, Firenze, Italy\\
\inst{3}Aix Marseille Universit\'e, Universit\'e de Toulon, CNRS, CPT UMR 7332, 13288 Marseille, France}

\abstract{
A one-dimensional long-range model of classical rotators with an extended degree of complexity, as compared to paradigmatic long-range systems, is introduced and studied. 
Working at constant density, in the thermodynamic limit one can prove the statistical equivalence with the Hamiltonian Mean Field model (HMF) and $\alpha$-HMF: a second order phase transition is indeed observed at the critical energy threshold $\varepsilon_c=0.75$. Conversely, when the thermodynamic limit is performed at infinite density (while keeping the length of the hosting interval $L$ constant), the critical energy $\varepsilon_c$ is modulated as a function of $L$.  At low energy,  a self-organized collective crystal phase is reported to emerge, which converges to a perfect crystal in the limit $\epsilon \rightarrow 0$. To analyze the phenomenon, the equilibrium one particle density function is analytically computed by maximizing the entropy. The transition and the associated critical energy between the gaseous and the crystal phase is computed. Molecular dynamics show that the crystal phase is apparently split into two distinct regimes, depending on the  the energy per particle $\varepsilon$. For small $\varepsilon$, particles are exactly located on the lattice sites; above an energy threshold $\varepsilon{*}$,  particles can travel from one site to another. However, $\varepsilon{*}$ does not signal a phase transition but reflects the finite time of observation:  the perfect crystal observed for $\varepsilon >0$  corresponds to a long lasting dynamical transient, whose life time increases when the $\varepsilon >0$ approaches zero.}


\maketitle

The study of long-range interacting systems experienced a renewed interest, in the last decades, due to the increasing computational resources made available on  modern computers. Improved numerical simulations allowed to enhance the statistics over previous investigations, enabling for new ideas to be tested and existing theory to be challenged \cite{review,book}. In a long-range system the two body potential decays with the distance $r$ as $r^{-\alpha}$, the exponent $\alpha$ being smaller than the dimension of the embedding space. Many different systems fall in such a wide category. Among others, it is worth mentioning self-gravitating systems and their applications to astrophysics and cosmology \cite{galaxy1,galaxy4}, charged plasma dynamics \cite{plasma1,plasma3}, hydrodynamics of two-dimensional vortices \cite{vortex1,vortex3}, spin-wave interactions \cite{spinwave1,spinwave2} with their implications for the physics of lasers \cite{fel}. 
As largely testified by recent advances, long-range systems display a rich zoology of peculiar behaviors, ranging from ensemble inequivalence to out-of equilibrium dynamics, and constitute an intriguing arena for novel developments, of both applied and fundamental interests. 

Several long-range models exist in the literature, that are in principle suitable for elucidating key aspects of both equilibrium and out-of-equilibrium dynamics.  Simplified toy models and paradigmatic case studies have been in particular proposed, which allow to progress in the analysis by 
imposing a substantial reduction in the inherent complexity of the inspected physical problems. Working along these lines, it is instructive to elaborate on the 
ingredients that drive the spontaneous emergence of coherent self-organized structures, as seen in real experiments.  As an example relevant for the topic addressed in this Letter, crystal-like structures have been reported to occur in long-range physical systems. as e.g. in dusty charged plasmas \cite{plasmacrist1, plasmacrist2}, Coulomb-interacting cold atoms \cite{coulombcrist} or Bose-Einstein condensates \cite{bosecrist}.  The phenomenology of these latter states is not clearly understood: experiments seem to suggest the presence of an intermediate phase during the melting of the solid structure, which displays additional ordering at different scales \cite{plasmacrist1}. When long-range interactions are active, the interplay between potential and kinetic energy terms give rise to correlations that do not decay rapidly with the distance. This latter effect may eventually translate into the emergence of non trivial thermodynamic transitions and peculiar symmetries. 

Available long range models often lack the necessary degree of phase space complexity to effectively reproduce the characteristics of such patterns. 
For instance, in the celebrated Hamiltonian Mean Field (HMF) model the spatial extension of the phase space is bounded in size, each particle's position being mapped onto the circle $\theta\in [0,2\pi[$. Furthermore, the notion of inter-particles' distance is not explicitly accounted for in the framework of the HMF formulation.
Other long-range solvable systems, albeit spatially extended, may lack of translational invariance, like the $\phi^4$ model \cite{review, book}, or encode for a peculiar lattice metric, as it is the case for the so called $\alpha$-HMF model \cite{review, book}. This latter has a predetermined skeleton that is embedded in the model itself, and which reflects in the observed structures as a natural imprint.\\

To bridge this gap and eventually describe the appearance of crystal-like long-range structures, we will here introduce and characterize a new model of coupled rotators, closely inspired to the aforementioned $\alpha$-HMF and HMF models. The model that we are going to introduce will spontaneously freeze into a crystal-like phase, at sufficiently low energies. When the energy gets increased, the self-organized crystal melts, resulting in a disordered gaseous medium.  
The Letter is organized as follows: we first briefly recall the thermodynamic properties of both the HMF and $\alpha$-HMF models. We then move to presenting the extended model, focusing on its equilibrium thermodynamical features, and demonstrating the existence of a crystal phase, via combined numerical and analytical means.

The HMF model has imposed itself in recent years as a paradigmatic model for the study of long range interacting systems \cite{hmf}. Indeed, it displays a rich variety of features, most notably related to its out-of-equilibrium dynamics. Moreover, computing its statistical properties via direct simulations of the microscopic dynamics is both simple and fast. The HMF model is characterized by the following Hamiltonian:
\begin{equation}
\label{hmf}
H=\sum_{i=1}^{N}\frac{p_{i}^{2}}{2}+\frac{1}{2N}\sum_{i,j=1}^{N}(1-\cos(q_{i}-q_{j}))\:,
\end{equation}
where $q_{i}$ and $p_{i}$ stand for the canonically conjugated variables of particle $i$, and $N$ denotes the total number of particles. The HMF model displays, at equilibrium, a second order phase transition at critical energy $\varepsilon_{c}=0.5$, with a characteristic order parameter defined as the global magnetization 
$\mathbf{M}=\frac{1}{N}\sum_{i}(\cos(q_{i}),\sin(q_{i}))$. The associated phase space is bounded: the positions of the particles can be mapped on the unitary circle, because of the periodicity of the potential and the mean-field nature of the interaction. In order to explicitly take into account the decay of the interaction potential as due to long-range couplings, 
an extension of the HMF model referred to as the $\alpha$-HMF model \cite{Anteneodo98,alpha1} has been proposed in the literature. In this generalized model, the rotators occupy the sites of a regular lattice and the Hamiltonian writes:
\begin{equation}
H=\sum_{i=1}^{N}\frac{p_{i}^{2}}{2}+\frac{1}{2\tilde{N}}\sum_{i,j=1}^{N}\frac{1-\cos(q_{i}-q_{j})}{||i-j||^{\alpha}}\:,
\label{alphahmf}
\end{equation}
where $\alpha$ is a free parameter, which enables one to adjust the strength of the interaction, and $\Vert i-j\Vert$ represents the distance between the lattice sites $i$ and $j$. The normalization factor reads $\tilde{N}=(2/N)^{\alpha}+2\sum_{i=1}^{N/2-1} (1/i^{\alpha})$ 
and ensures extensiveness. For finite sizes, in order to keep the system statistically invariant by translation along the lattice, it is convenient to impose periodic boundary condition. To accomplish this, it is customary to confine the lattice on a circle of radius $(N-1)/2\pi$ and then consider in (\ref{alphahmf}) the minimum distance on the circle between any pair of selected sites.  For $\alpha<1$, the $\alpha$-HMF is thermodynamically equivalent to the HMF model 
\cite{Campa2000,alpha1,Vandenberg10,alpha2,alphatrans}, albeit it displays a lattice ordering enforced by the model in its microscopic formulation.\\

Starting from these premises, and to eventually explore the possibility of generating a non trivial crystal-like phase, which is not the mere byproduct of the underlying discrete spatial support, we here propose a straightforward generalization of the aforementioned $\alpha$-HMF model, that we will refer to as to the $\beta$-HMF.  Particles are no longer forced to sit on specific lattice sites, but can freely explore the hosting support. In practice, we substitute in Hamiltonian (\ref{alphahmf}) the term $\Vert i-j\Vert$ by $\Vert q_{i}-q_{j}\Vert$, a replacement which  does not produce divergences in the force field, as long as $\alpha<1$. Introducing a continuum distance parameter, allows in turn to extend the phase space that particles can explore while, at the same time, removing the artificial constraint of a fixed lattice support. 
The $\beta$-HMF model is in some respects reminiscent of the self-gravitating ring model \cite{numscheme}, but, as opposed to this latter, it does not require the introduction of an artificial regularization of the potential at small scales. In this Letter, we will always deal with periodic boundary conditions on a circle of length $L$. Then, we will consistently define the distance between two particles with position $q_i$ and $q_j$ as $d_{ij}=\min\lbrace |q_i-q_j|,L-|q_i-q_j|\rbrace := \Vert q_i - q_j \Vert$
namely, the shortest arc-length on the circle. Based on the above,
the Hamiltonian of the $\beta$-HMF can be cast in the form \cite{alphahmffourier}:
\begin{equation}
\label{betahmf}
H=\sum_{i=1}^{N}\frac{p_{i}^{2}}{2}+\frac{A(N,L)}{2}\sum_{i,j=1}^{N}\frac{1-\cos(q_{i}-q_{j})}{||q_{i}-q_{j}||^{\alpha}}
\end{equation}

When $0\leq\alpha<1$ the system is formally long-range: this latter parameter can  be tuned to control the strength of the long-range interaction. $A(N,L)$ is a normalization constant which guarantees extensiveness, following the Kac prescription \cite{kac}, and reads
 $A=\frac{(1-\alpha)}{N}\left(\frac{L}{2}\right)^{\alpha}$.
The above relation can be rigorously derived in the large scale limit and holds in general, as confirmed by numerical simulation. 
As a consistency argument, we remark that the expected normalization factor are recovered both in 
the HMF ($\alpha=0$) and $\alpha-$HMF ($q_{i}=2\pi k+\xi_{i}$, for $L\to\infty$, $\xi_{i} \in [0,2\pi[$ and $k \in [0,L[$) limits. 
As a further, preliminary comment, we note that as $q_{i}$ are assigned to label the  particles' positions, in the interaction potential one should 
write in principle $\cos(z(q_{i}-q_{j}))$, where $z$ is a suitable constant that makes the argument of the cosine adimensional. Without loss of generality, we set $z=1$ in the following. Moreover, in this Letter, we will assume $L=2\pi l$, $l\in\mathbb{N}$, and introduce the rescaled density  $\rho_0=N/l$ to simplify the notation.\\

We start our analysis by focusing on the low temperature regime. In this limit we expect to obtain minimal energy states. From the two-body interaction potential, 
one can imagine that the system will self-organize so to fill distinct available minima. It is however not evident that the process should eventually materialize in a crystal like structure,  condensation of all particles in just one minimum being a priori also possible. 
Indeed, the further apart the particles sit, the wider the potential minimum appears. Hence, for a fixed energy amount, particles are more prone to fluctuate if they populate distant sites, so endowing a global repulsive entropic force. 
Since we assume particles to populate a bounded circular domain, we expect that the entropic repulsion will oppose condensation, so favoring in turn the spontaneous emergence of the crystal state, at low temperature.\\

To test the adequacy of this interpretative scenario, we carried out a campaign of $N$-body simulations. These are performed by using a fifth-order symplectic scheme \cite{Atela}, with a time-step $\delta t=0.05$. The equation of motion can be derived from Hamiltonian (\ref{betahmf}). 
We wish to emphasize that the explicit presence of the inter-particle distance in (\ref{betahmf}) heavily increases the required computational cost, as compared to the simple HMF case study, or even the $\alpha$-HMF model \cite{alphahmffourier}). For this reason, our analysis is limited to a maximum of $N=10^{3}$ particles. The time needed to converge to equilibrium gets clearly larger, as the population size is increased. As an additional remark, we notice that for $N=2$ the system is integrable. In this limiting case, one can analytically access the phase space of the system, and use the information to test the numerical implementation. It is indeed extremely useful to benchmark numerical and analytical outcomes to avoid technical pitfalls which might originate when the subtle condition $||q_{i}-q_{j}||\sim 0$, is dynamically met. Furthermore both total energy and momentum are constants of motion, due to the translational invariance, and were numerically monitored over time. All simulations presented in this Letter assume an initial Gaussian distribution of momenta, with total momentum $P=0$. For any given value of the energy, the system is observed to converge to the same equilibrium state, independently of the distribution positions $q$, assigned at $t=0$: we tested $q_i=2\pi k$, $q_i=0$, $q_i$ from a uniform distribution, or Gaussian centered either $q=0$ or on each $q=2\pi k$, which was the fastest to converge.\\
To characterize the thermodynamic fate of the system one can in principle rely on the magnetization $M=|\mathbf{M}|$. When particles are uniformly spread over the hosting domain, we have $M=0$. Conversely, $M=1$ is equally associated to bunching on a single cluster, the condensed phase to which we alluded above, or to a distribution peaked on sites which are $2\pi$-periodic, namely the sought crystal state. To remove this degenerancy and discriminate between a crystal ordering phase and a condensate solution with all particles close to each other in space,  one needs to compute the distribution of relative particles' positions.\\
\begin{figure}
\begin{centering}
\includegraphics[width=0.45\textwidth]{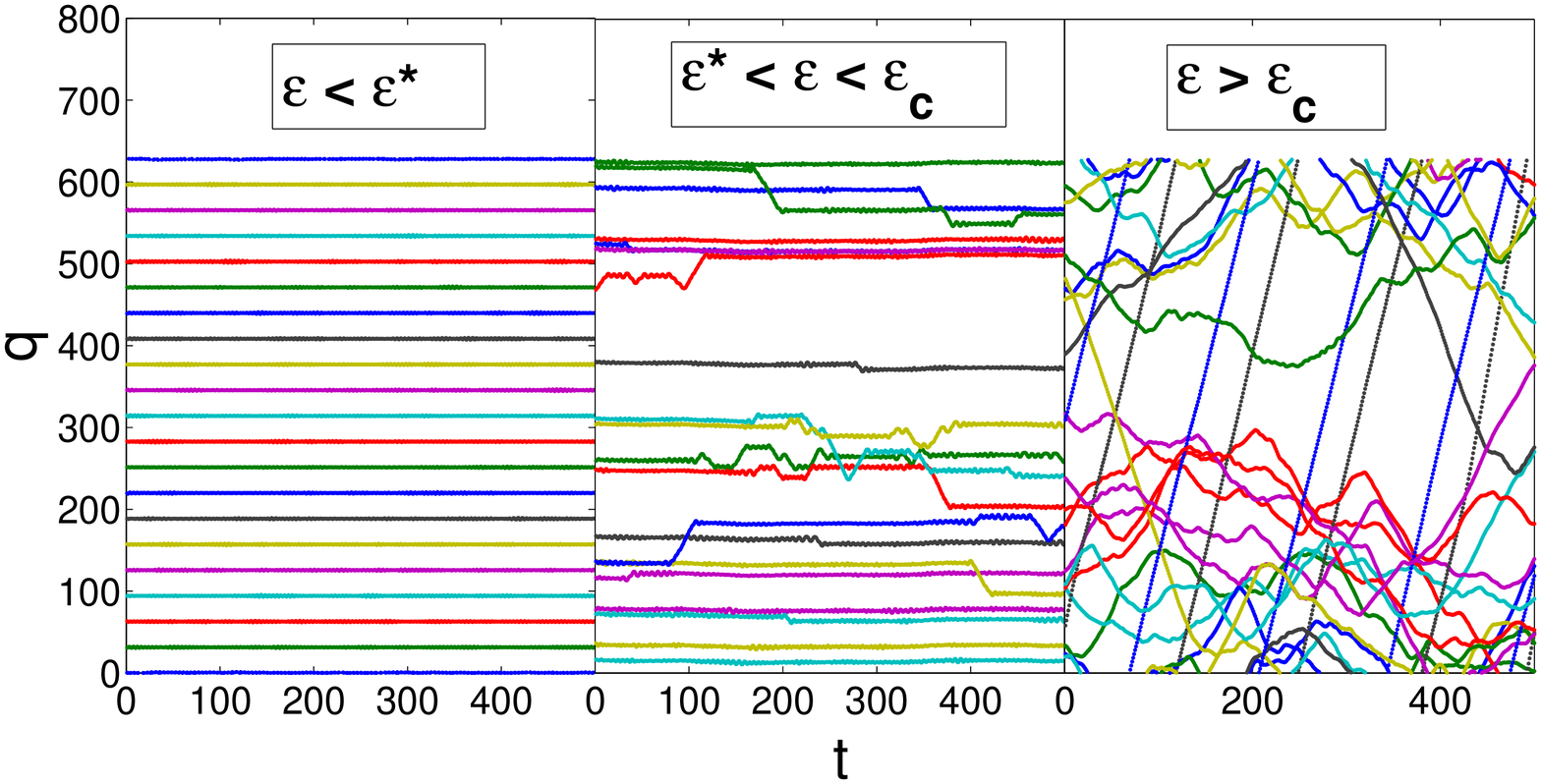}
\end{centering}
\caption{\footnotesize A few tracer particles positions in the q-space as function of time. The analysis of the recorded trajectories shows that a priori three dynamical regimes exist for  the $\beta$-HMF model (\ref{betahmf}), depending on the energy per particle $\varepsilon$. At $\varepsilon=\varepsilon{*}\sim0.1$, the perfect crystal like structure is lost, and the particles can wonder to neighboring sites. Above $\varepsilon_c\sim0.75$ the self-organized crystal melts, resulting in a disordered gaseous medium. Here $N=500$, $L=N$ and $\alpha=0.5$.}
\label{phaseline}
\end{figure}
We first start by qualitatively looking at some particle trajectories, $q$ versus time, $t$, as obtained through direct $N$-body simulations. Those are displayed in Fig.~\ref{phaseline}. Three rather different equilibrium regimes are apparently detected. At very low energy densities, the system self-organizes into a crystal in which particles are bounded in the vicinity of what appear to be lattice sites. When increasing the energy per particle above a threshold amount that we termed
$\varepsilon{*}\sim0.1$, the system enters a partially disordered phase: particles are nevertheless localized over finite windows of time, and the global magnetization is still positive; a ``soft'' crystal phase emerges. 
At higher energies, for $\varepsilon>\varepsilon_{c}\sim0.75$, the system loses any traces of regularity and the crystal melts into a gaseous medium. As we shall prove in the following, only two equilibrium regimes exist: a self-organized crystal with no particle-site correspondence and a gaseous disordered phase. In other words, $\varepsilon{*}$ does not signal a genuine transition but indirectly stems from the finite time of observation assumed in the simulations, as in the spirit of the Arrhenius law. The perfect crystal exists only in the limit $\varepsilon \rightarrow 0$, when the energy barrier between two adjacent site diverges: the perfect crystal displayed for $0<\varepsilon<\varepsilon{*} $  corresponds to a long lasting dynamical transient, whose life time increases when the zero energy limit is eventually approached.

To cast this observation on solid grounds, we calculated analytically the single particle distribution as follows a maximization of the thermodynamic entropy, subject to the constraints of the dynamics. As compared to the HMF and $\alpha$-HMF models, we have one more degree of freedom in choosing the thermodynamic limit. The system now depends on both the rescaled length $l$ and the average particle density $\rho_0$. When performing the limit $N\to\infty$ we can consider $\rho_0\to\infty$ at fixed $l$, or, conversely, the limit $l\to\infty$ at constant $\rho_0$. Indeed, the dependence on both  $\rho_0$ or $L$ can be made explicit in the scaling constant $A$. 
In the dual limit, when the density is kept constant and $l\to\infty$, it is not so straightforward to analytically  recover the analogue Hamiltonian. The problem lies in that the potential is modulated by the cosine with a periodicity of $2\pi$, but it is not periodic. So we are not allowed to perform trivial simplifications without any knowledge on the form of the distribution function. 
Our strategy will be to solve the system for finite length, and observe the scaling of the associated thermodynamic quantities for $l\to\infty$. 

To tackle the infinite density limit ($N\to\infty$ with $l$ constant), we consider the one-particle density function $f(q,p,t)$ and denote the particle density as $\rho(q,t)=\int f(p,q,t)dp$.
$f(q,p,t)dqdp$ represents the fraction of particles in the interval $[q, q+dq][p,p+dp]$ at time $t$. The system conserves the energy density 

\begin{equation}
\label{limitH}
\lim_{N\rightarrow\infty}\frac{H}{N}=\varepsilon[f]=\int_\Gamma f(q,p,t) \left(\frac{p^2}{2} + V(q,t)\right) dq\:dp\: ,
\end{equation}

where $\Gamma$ represents the phase space volume, and 
\begin{equation}
\label{Vq}
V(q,t)=  \int \rho(q^\prime,t) \frac{1-\alpha}{2\pi L}\left(\frac{L}{2} \right)^{\alpha} \frac{1-cos(q-q^\prime)}{||q-q^\prime||^\alpha} \, dq^\prime \: .
\end{equation}

Also note that due to translational invariance, the total momentum $P[f]=\int_\Gamma f(q,p,t)p dq dp$ is conserved together with normalization condition $\mathcal{G}[f]=\int_\Gamma f(q,p,t) dq dp = 1$. 
We assume then a strictly Boltzmannian entropy $S[f]= - \int_\Gamma f(q,p,t) \log(f(q,p,t)) dq dp$. 
To obtain the equilibrium function $f(q,p)$ which maximizes the entropy $S[f]$, under the constrains of the dynamics imposed, we have to solve the maximization problem $\max_f\lbrace S[f] | \int_\Gamma f(q,p,t) dq dp = 1, \varepsilon[f]=E, P[f]=P \rbrace$. By introducing the Lagrange multipliers $\beta$, $\gamma$ and $\mu$, 
and requiring stationarity of the entropy functional, one readily obtains:
\begin{equation}
\label{distrfun}
f(q,p) = D e^{-\beta(\frac{p^2}{2} + V(q))} \: ,
\end{equation}
where and $D=e^{-1 -\mu}$ and the equilibrium condition leaves $V$ and $f$ time independent.
In the above calculation we made use of $\gamma=0$, as $P[f]=0$, for our class of initial conditions. Expression (\ref{distrfun}) is the standard Maxwellian distribution, where $\beta$ is the inverse temperature and $\mu$ stands for the chemical potential. As for the HMF model, one obtains an implicit system, since $V(q)$ depends in turn on $f(q,p)$. To solve the problem we use an iterative numerical scheme introduced in \cite{numscheme}, which generalizes the standard Newton technique by ensuring that entropy steadily increases towards the maximum value at each iteration step.
From equation (\ref{distrfun}) we find that the equilibrium density function $\rho(q)$ becomes: $\rho(q) = \int f(q,p) dp = D\sqrt{\frac{2\pi}{\beta}}e^{-\beta V(q)}$ \:.

\begin{figure}[h!]
\centering
\includegraphics[width=0.45\textwidth]{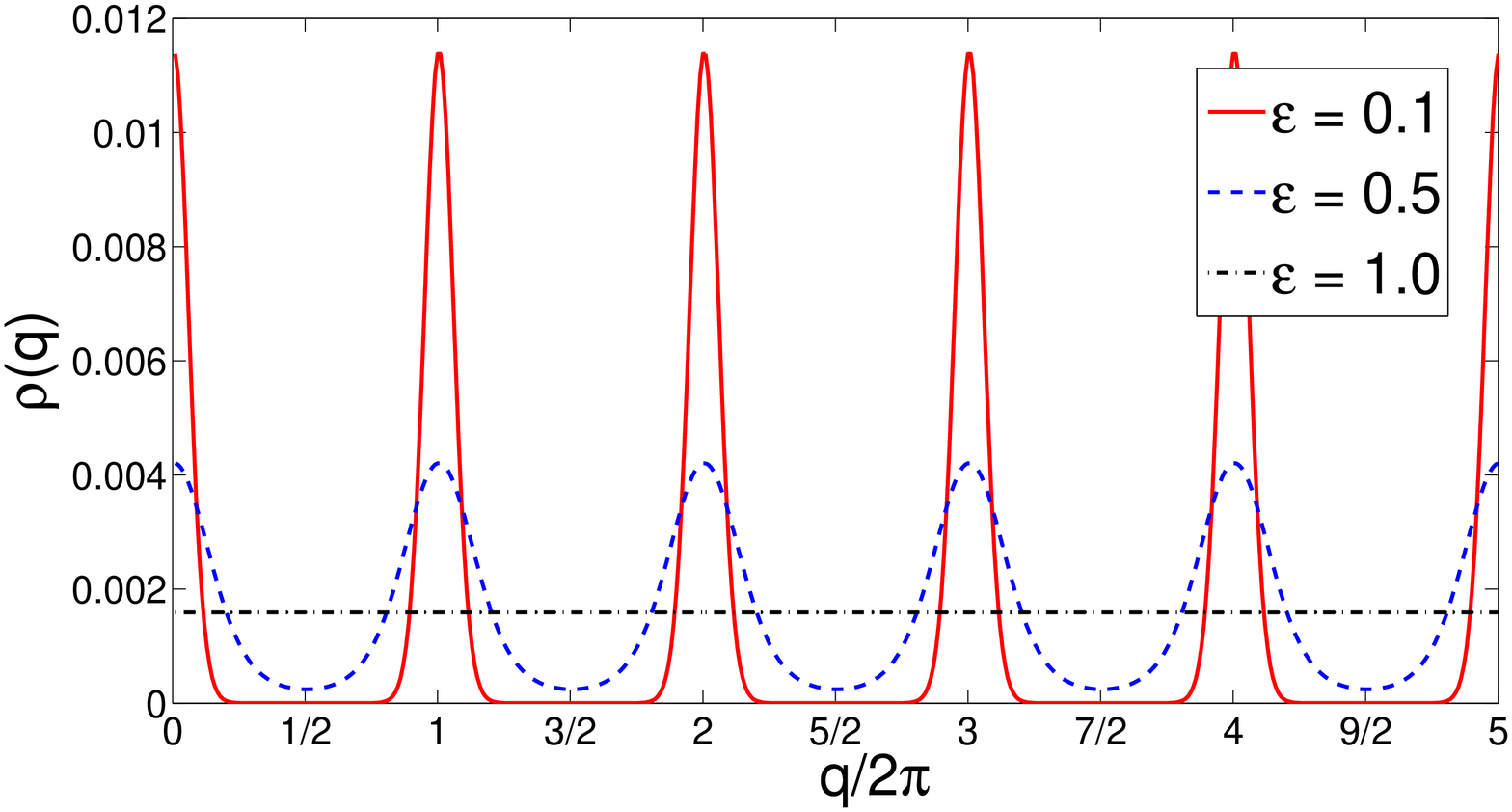}
\caption{\footnotesize Analytically predicted equilibrium distribution densities $\rho(q)$ for different dynamical regimes and $\alpha=0.5$. Here we plot a small portion of the $q$ axis with 5 periods for a distribution function computed for $l=100$ periods. The crystal lattice appears as expected.}
\end{figure}

\begin{figure}
\centering
\includegraphics[clip,width=0.45\textwidth]{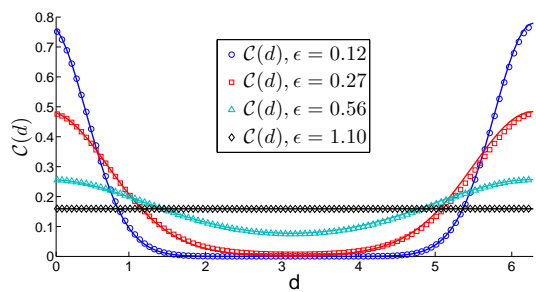}
\caption{\footnotesize The theoretical distribution $\mathcal{C}(d)$ of inter-particle distances for infinite density (solid lines)  is compared to direct simulations for  $\rho_0=1$ (symbols). Here $l=500$ and $\alpha=0.5$. Different curves correspond to distinct energy densities $\varepsilon$. The horizontal axis is rescaled in a $\mod(2\pi)$ representation and the curves are normalized so to have unitary area. $\mathcal{C}(d)$ is translationally invariant: an almost perfect matching is observed.\label{rhofun02}}
\end{figure}

A comparison between the theoretically predicted $\rho(q)$ and the result of numerical simulations can be easily drawn for $\varepsilon<\varepsilon_*$, when particles occupy  the self-emerging lattice, and for $\varepsilon>\varepsilon_c$, when particles are homogeneously spread. In both cases, we observe a good matching between theory and simulations, even with a low number of particles  and low densities (data not shown). In the intermediate phase $\varepsilon_*<\varepsilon<\varepsilon_c$, averaging effects due to particles relocations makes the direct comparison more cumbersome, since one has to cope with the the fact that due to translational invariance we have as well local vibrations of the crystal.
To overcome these limitations, in figure \ref{rhofun02} we confront the analytical predictions for the distribution of the inter-particle distances $\mathcal{C}(d) = \int \rho(q + d)\rho(q) \, dq$ to the simulated data. $\mathcal{C}(d)$ is by definition translational invariant and the agreement between theory and simulations is almost perfect, for all the explored energy regimes.\\

\begin{figure}
\centering
\includegraphics[clip,width=0.45\textwidth]{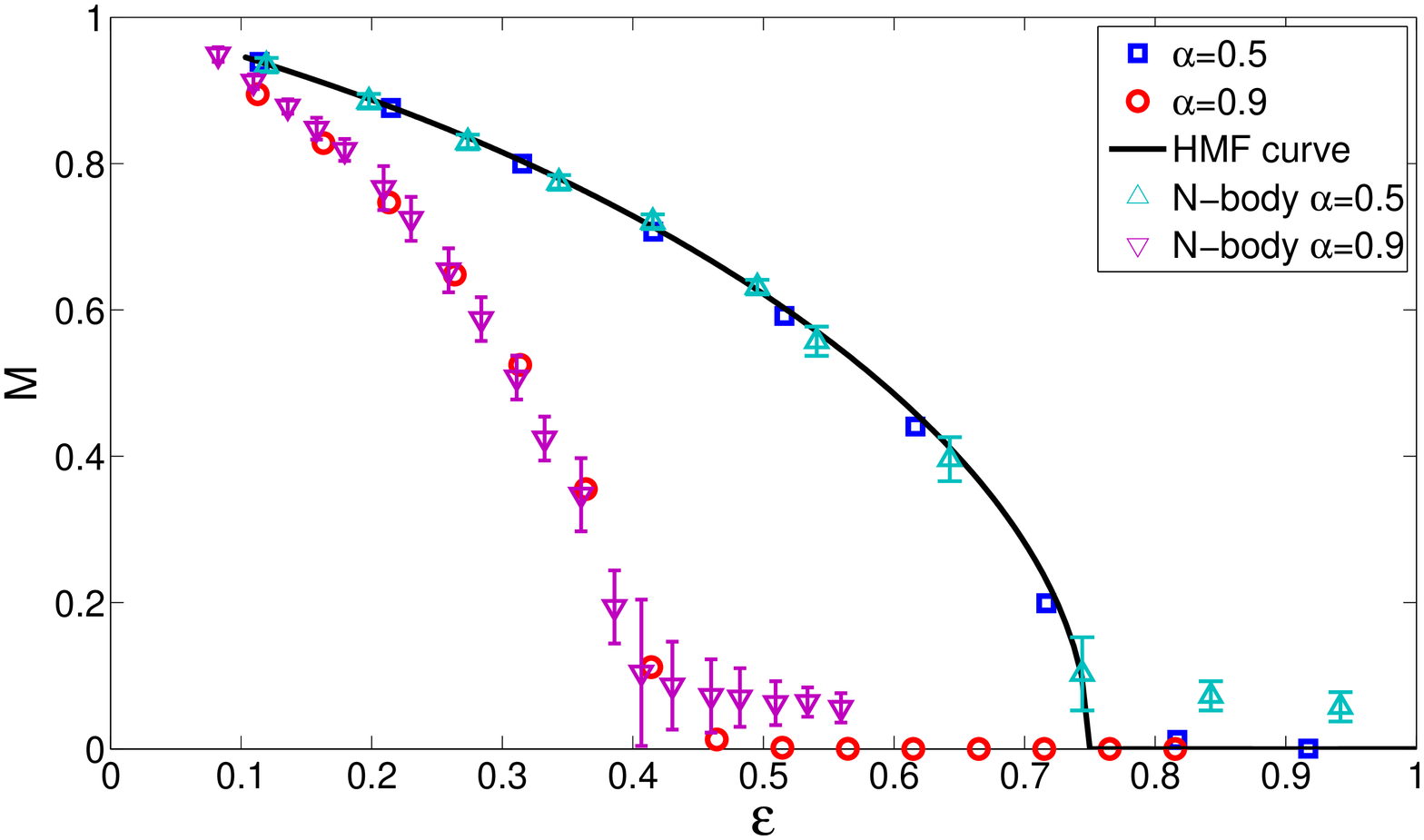}
\caption{\footnotesize Comparison between the analytic magnetization curves and the data recorded via $N$-body simulations, for a random flat initial condition over the space positions. Data refers to $l=1000$. An average over a time window $\Delta t=5000$ is performed. As expected data is noisy close to the critical transition energy $\varepsilon_{c}$. The black thick curve represents the HMF equilibrium curve. \label{Mphas2}}
\end{figure}

As an additional test, we proceed to compute the curve for the equilibrium magnetization $M(\varepsilon)$, for different values of the scaling exponent $\alpha$ and for $l=1000$. A second order phase transition between ferromagnetic $M>0$ and paramagnetic $M=0$ phases is found to occur at a  critical energy density $\varepsilon_c$, see Figure \ref{Mphas2}. Remarkably, for $\alpha<0.6$ (for the chosen value of $l$) the data collapse on the HMF transition curve and $\varepsilon_c = 0.75$. For larger values of $\alpha$ the curves shifts towards the left and, consequently, $\varepsilon_c$ gets reduced.

As anticipated the numerically determined energy threshold $\varepsilon*$
which putatively separates the perfect and soft crystals,
does not leave a trace in the magnetization curves.
For both large and small values of the scaling exponent $\alpha$ the agreement between theory and simulations is convincing. A dedicated campaign of investigations suggests that, for all values of $\alpha$, the transition energy $\varepsilon_c$ converges to the asymptotic value $0.75$, in the limit $l \to \infty$. In this limit, the transition curves of the $\beta$-HMF and HMF (and clearly $\alpha$-HMF) superpose to each other. In the following we will provide an independent argument to show that the $\beta$-HMF reduces to the $\alpha$-HMF, when $l \to \infty$. 

Let us start by simplifying the expression of the distance term which appears in Hamiltonian (\ref{betahmf}). Define $q_{j}=2\pi h_{j}+\delta_{j}$, with $h_{j}\in\{0,\cdots,n\}$ playing the role of a lattice index and $0<\delta_{j}<2\pi$. Given the equilibrium distribution $\rho(\delta,h)$ computed above,
in the limit $L\to\infty$, we ignore the contribution to the potential coming from pair of particles $i$ and $j$ which are close and have $h_i=h_j$. Under these working assumptions, for $h_i \ne h_j$ the distance term in equation (\ref{betahmf}) can be simplified as:
\begin{equation}
\label{scale2}
\begin{split}
& \frac{1}{\left\Vert \frac{2\pi h_{i}}{L}-\frac{2\pi h_{j}}{L}+(\delta_{i}-\delta_{j})/L\right\Vert ^{\alpha}} \approx \\
& \approx \frac{1}{\left\Vert \frac{2\pi h_{i}}{L}-\frac{2\pi h_{j}}{L}\right\Vert ^{\alpha}}-\alpha\frac{(\delta_{i}-\delta_{j})}{L\left\Vert \frac{2\pi h_{i}}{L}-\frac{2\pi h_{j}}{L}\right\Vert ^{\alpha+1}} \: .
\end{split}
\end{equation}
The second term in Eq. (\ref{scale2}) expression vanishes, when performing the sum in the limit $L\to\infty$. One loses therefore the dependence on $\delta$. In other words, the numerator and denominator of the potential in equation (\ref{betahmf}) become independent and the potential scales as 
$\sum_{j=1}^N ({1-\cos(\delta_i-\delta_j)})/({\left\Vert h_{i} - h_{j} \right\Vert ^{\alpha}})$. Since the equilibrium distribution function is homogeneous in the index $h$ and since  each interval (of length $2\pi$) in which we have sub-divided the system, contains the same number of particles, we can reorder the sum in the potential by avoiding to  individually count the particles which share the same index $h_i$. We then perform the replacement $h_i \to i$. Apart for a multiplying constant depending on $\rho_0$, we recover the form of the $\alpha$-HMF potential (\ref{alphahmf}). Since the steps are independent of $\alpha$, we speculated that the $\alpha$-HMF equilibrium solution, equivalent to  the original HMF model, is also found within the $\beta-$ HMF setting, for any value $0 \leq \alpha < 1$, provided $l$ is sufficiently large. 

We now return to Fig.~\ref{rhofun02} to discuss again the long-lasting dynamical regime of perfect crystal. Particles are confined on lattice sites, as enforced by the fact that $\rho(q)$ displays a periodic set of minima with $\rho\sim 0$. For high enough energies the minimum $\rho_{min}$ of $\rho(q)$ lifts clearly above zero: even though the crystal seems to preserve a well defined structure, this latter reflects a collective behaviour. Fluctuations instigate an effective mixing between adjacent lattice sites. If a truly transition between perfect and ``soft'' crystal existed, the gap between consecutive sites should close $\rho_{min}=0$ for some value of the energy density, so resulting in an infinite barrier for the particles to cross as in the spirit of the Arrhenius law.    
Due to the observed analogy between the $\beta$-HMF and $\alpha$-HMF models, the distribution of densities ($q$ modulo $2\pi$) obtained for the $\beta$-HMF model superposes exactly (data not shown) to those obtained for the HMF.
The interest of this conclusion are twofolds. On the one side, this observation implies that the universality class of the HMF extends to embrace non trivial models, as the $\beta$-HMF is. Then, more importantly for the aforementioned issue,  $\rho_{min}\sim\exp(-\beta M)$ for $q=\pi$, meaning that $\rho_{min}$ formally vanishes only at zero temperature. Hence, as anticipated, only at zero energy density (zero temperature), one finds a perfect crystal lattice, particles being {\it de facto} trapped on specific lattice sites. For a non zero, but still small, energy density, we end up with a  collective crystal structure with particles that can in principle move from one site to another, while preserving a macroscopic crystal-like self-organization. The lattice emerges as a collective property of the distribution, and the trapping  is effective for finite, though potentially large, times. The observed energy threshold $\varepsilon*$ separating perfect and ``soft'' crystal regimes reflects therefore the finite time of observation and the finite size of the system rather then representing a genuine transition between two distinct equilibrium states.\\

Summing up, we have here introduced and studied a novel long-range model, to shed further light on  
those spatial self-organized structures that can emerge when long-range coupling are at play. A continuous inter-particles distance term adds a degree of complexity with respect to the 
HMF and $\alpha$-HMF models, while reproducing similar thermodynamics behaviours. In the extended phase space of the $\beta$-HMF it is possible to observe a more rich phenomenology of equilibrium features.
Specifically, we unravelled two different states,  a self-organized effective crystal phase and a gaseous phase at high energies, which could not be seen in the reduced phase space of previous long-range models. The crystal phases manifests as a generalized self-organized state, in which the local lattice emerge from the collective behaviour of the particles. A solid to gas transition manisfests at a critical energy which is sensitive to the scaling size $L$, and converges to 0.75 in the limit $L \to \infty$.
In the crystal phase, a distinction between ``perfect'' and collective crystal can be observed for finite sizes and finite time. Moreover we were able to analytically obtain the equilibrium distribution function and apply it to describe the observed system behavior. Assuming the validity of the latter solution, the system can be formally reduced to the HMF in the limit $l \to \infty$, thus reproducing the same equilibrium transition between magnetized and homogeneous states. When operating at constant $l$, but infinite density, the $\beta$-HMF retuns a different equilibrium transition curve as compared to the HMF model.  Further studies should aim at evaluating the  interest of the crystal phase predicted within the realm $\beta-$HMF model, in light of its potential applications to real long-range systems of experimental relevance. For instance the fact there is no direct correspondence between emerging crystal sites and effective particle position could be reminiscent of supersolid properties. Another open question has to do with the possible presence of quasi-stationary out-of-equilibrium states \cite{vlasovlim,Vandenberg10,hmf3} which were non investigated in this paper.

\begin{acknowledgments}
X. L. thanks M. Adda-Bedia and T. L. Van den Berg for useful discussions during the setting up of the $\beta-$HMF model. X. L. is partially supported by
the FET project Multiplex 317532.
\end{acknowledgments}


\begin{thebibliography}{99}
\bibitem{review} A.~Campa, T.~Dauxois, S.~Ruffo, Physics Reports \textbf{480}, 57-159 (2009).
\bibitem{book} A.~Campa, T.~Dauxois, D.~Fanelli, S.~Ruffo, Physics of Long Range Interacting Systems, Oxford University Press (2015).
\bibitem{galaxy1} D.~Lynden-Bell, R.~Wood, Mon. Not. R. Astron. Soc.\textbf{138} 495 (1968).
\bibitem{galaxy4} P.~H.~Chavanis, I.~Ispolatov, Phys. Rev. E \textbf{66}, 036109 (2002).
\bibitem{plasma1} M.~K.~H.~Kiessling and T.~Neukirch, Proc. Nat. Acad. Sci. USA \textbf{100}, 1510 (2003).
\bibitem{plasma3} P.~H.~Chavanis, Eur. Phys. J. B \textbf{52}, 61 (2006).
\bibitem{vortex1} R.~A.~Smith and T.~M.~O'Neil, Phys. Fluid. B \textbf{2}, 2961 (1990).
\bibitem{vortex3} R.~S.~Ellis, K.~Haven, and B.~Turkington, Nonlinearity \textbf{15}, 239 (2002).
\bibitem{spinwave1} J.~Barr\'e, D.~Mukamel, S.~Ruffo, Phys. Rev. Lett. 87, 030601 (2001).
\bibitem{spinwave2} D.~Mukamel, S.~Ruffo and N.~Schreiber, Phys. Rev. Lett. 95, 240604 (2005).
\bibitem{fel} J.~Barr\'e T.~Dauxois, G.~de~Ninno, D.~Fanelli, S. Ruffo, Phys. Rev E \textbf{69}, 045501(R) (2004).
\bibitem{plasmacrist1} H.~M.~Thomas, G.~E.~Morfill, Nature \textbf{379}, 806 - 809 (1996).
\bibitem{plasmacrist2} M.~H.~Thoma, M.~Kretschmer, H.~Rothermel, H.~M.~Thomas, G.~E.~Morfill, Am. J. Phys.\textbf{73}, 5 (2005).
\bibitem{coulombcrist} A.~Bermudex, J.~Almeida, F.~Schmidt-Kaler, A.~Retzker, M.~B.~Plenio, Phys. Rev. Lett., 107 (2011) 207209, arXiv:1108.1024v2 (2011).
\bibitem{bosecrist} Kazimierz~Lakomy, Rejish~Nath, Luis~Santos, Phys. Rev. A, 85 (2012) 033618(2011).
arXiv:1107.3132v2 (2011).
\bibitem{hmf} M.~Antoni and S.~Ruffo, Phys. Rev. E \textbf{52}, 2361 (1995).
\bibitem{Anteneodo98} C.~Anteneodo, C.~Tsallis, Phys. Rev. Lett. \textbf{80}, 5313 (1998).
\bibitem{Campa2000} A.~Campa, A.~Giansanti, and D.~Moroni, Phys. Rev. E \textbf{62}, 303 (2000).
\bibitem{alpha1} F.~Tamarit, C.~Anteneodo, Phys. Rev. Lett. \textbf{84}, 208 (2000).
\bibitem{Vandenberg10} T.~L.~{Van den Berg}, D.~Fanelli, X.~Leoncini, EPL \textbf{89}, 50010 (2010).
\bibitem{alpha2} R.~Bachelard, T.~Dauxois, G.~De~Ninno, S.~Ruffo, F.~Staniscia, Phys. Rev. E \textbf{83}, 061132 (2011).
\bibitem{alphahmffourier} T.~L.~{Van den Berg}, \textit{Systèmes avec interaction à longue portée}, Master thesis, director X. Leoncini (2009).

\bibitem{kac} M.~Kac, G.~E.~Uhlenbeck, P.~C.~Hemmer, J. Math. Phys. \textbf{4}, 216 (1963).

\bibitem{Atela} R.~I.~McLachlan, P.~Atela, The accuracy of symplectic integrators, Nonlinearity \textbf{5}, 541-562 (1992).
\bibitem{vlasovlim} A.~Antoniazzi, F.~Califano, D.~Fanelli, S.~Ruffo, Phys. Rev. Lett. 98, 150602 (2007).
\bibitem{alphamot} X.~Leoncini, T.~L.~Van den Berg, D.~Fanelli, EPL \textbf{86} 20002 (2009).
\bibitem{hmf3} A.~Antoniazzi, D.~Fanelli, J.~Barr\'e, P.~H.~Chavanis, T.~Dauxois, S.~Ruffo, Phys. Rev. E \textbf{75}, 011112 (2007).
\bibitem{numscheme} T.~Tatekawa, F.~Bouchet, T.~Dauxois, S.~Ruffo, Phys. Rev. E \textbf{71}, 056111 (2005).
\bibitem{alphatrans} A. Turchi, D. Fanelli, X. Leoncini, Comm. Nonlin. Sci. Num. Sim. \textbf{16}, 4718-4724 (2011).


\end{thebibliography}
\end{document}